\documentclass[aps,prl,twocolumn,groupedaddress,showpacs]{revtex4}

\usepackage{graphicx}
\begin{document}


\title{Snell's Law for Shear Zone Refraction in Granular Materials}


\author{H. A. Knudsen}
\email[]{h.a.knudsen@fys.uio.no}
\author{J. Bergli}
\email[]{jbergli@fys.uio.no}
\affiliation{Department of Physics, University of Oslo, P.O. Box 1048 Blindern, NO-0316 Oslo, Norway}

\date{\today}

\begin{abstract}
We present experiments on slow shear flow in a split-bottom linear shear
cell, filled with layered granular materials. Shearing through two
different materials separated by a flat material boundary is shown to give
narrow shear zones, which refract at the material boundary in accordance
with Snell's law in optics. The shear zone is the one that minimizes the
dissipation rate upon shearing, i.~e.~a manifestation of the principle of
least dissipation. We have prepared the materials as to form a granular
lens. Shearing through the lens is shown to give a very broad shear zone,
which corresponds to fulfilling Snell's law for a continuous range of paths
through the cell.
\end{abstract}


\maketitle

Shearing, deformation or processing of granular materials are of practical
concern on many scales: faults and land slides in geological systems,
industrial powder processing, or mixing of materials. Flows occur in loose
to dense granular systems; fast or slow\cite{JNB96}.
Flow in dense systems tend to be
localized to narrow regions, shear bands, whereas most of the material is
moved as solid blocks. Shear bands may be down to a few particles
thick\cite{mueth2000}, but
wide zones and bulk flow also occur, very much depending on the boundary
conditions of the flows\cite{fenistein03,fenistein04,fenistein06}.

From a physical viewpoint, in sheared or processed granular materials
energy is dissipated. They will come to rest unless driven, be it by
gravity, by moving walls or by any other means. In many cases a
steady-state is reached, where the dissipated energy balances the input
energy. In classical mechanics the time evolution of any
system may be deduced from the principle of least action, and in optics the
corresponding principle of least time implies Snell's refraction law for
light.  A corresponding extremum principle for dissipative processes has
been sought after and \emph{the principle of least dissipation} can account
for the dynamics of some systems, but not for others\cite{onsager1931,ball1999}.
Shearing of granular materials is a process for which the principle is
proposed to be valid\cite{unger04}.
In a modified split-bottom Couette-cell setup the granular material
was forced into a steady-state with a localized shear
zone in the bulk of the material\cite{fenistein03,fenistein04,fenistein06}.
This was done with a single material, and by applying a minimum dissipation
principle the shape of the shear zone could be
explained\cite{unger04}. This principle must \emph{in principle} take the
form of a volume integral, namely that the motion is such that the
dissipation rate integral is minimized. We consider the system to be
invariant in the $z$-direction, in which case we propose the following form
of the integral
\begin{equation}
\label{eq:volint}
  \int p\mu \left| \nabla v(x,y) \right| dx dy dz = {\rm min}\ ,
\end{equation}
where $v(x,y)$ is the $z$-direction velocity field. The shear zone is here
considered a continuous sequence of thin surfaces with face normal $\nabla v$. 
The dissipation rate is found by multiplying with the shear force between
two sliding planes: the overall pressure $p$ times the effective friction
coefficient $\mu$.
For now, we make
the simplification that shearing is localized to narrow shear bands, in
which most of the sliding and thus energy dissipation take place. This
allows the reduction of the integral to be over the surface of the
shear zone, and we recover the form used in \cite{unger04,unger07}
\begin{equation}
\label{eq:surfint}
  \int \Delta v p \mu dS = {\rm min}\ .
\end{equation}
Here $\Delta v$ is the velocity difference between the two sides.  A
particular manifestation of the minimization principle is that a shear zone
going through a layer of one material into a layer of a different material
will be refracted at the material boundary.  Numerical work with idealized
particles and periodic boundary conditions suggested that the effective
friction $\mu$ may be understood as a \emph{material index}\cite{unger07},
from which it followed that the refraction of the shear zone follows
Snell's refraction law.

\begin{figure}[]
\includegraphics[width=0.95\columnwidth]{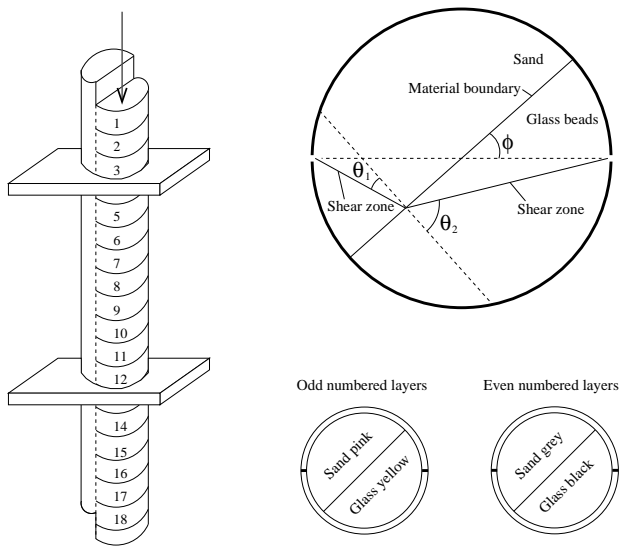}
\caption{\label{fig:setup}
Sketch of the experimental setup. To the left: the geometry of the shear
cell. Upper right: top view of a cut through the cell. Lower right: the
filling of material makes color layers.
}
\end{figure}

We test this experimentally by the setup sketched in
Figure \ref{fig:setup}. A purpose-built split-bottom shear cell is filled
with two materials, ``coarse'' sand and ``smooth'' glass beads. The
material boundary can be prepared at different angles with respect to the
shear cell.
The shear cell is a Plexiglas tube of length 37 cm,
outer radius 2 cm, and side thickness 2 mm.  It is cut in two parts which
slide with respect to each-other; mounted vertically through holes in two
wooden boards. The inside walls are roughened by horizontal stripes of 2 mm
glass beads glued on.  The cell is filled vertically half and half with
sand and glass beads. Upon filling a thin plastic divider is placed within
the tube at an angle $\phi$ with respect to the direct cut through the
cell.  After every 2 cm of filling it is raised correspondingly in order to
allow the material to settle and create contact over the material
boundary. The ``coarse'' material is sand; between 50 $\mu$m - 1 mm, highly
non-spherical. Pinkish sand is obtained by adding a tiny amount of color
powder. The ``smooth'' material is fabricated glass beads in two colors;
between 50 to 200 $\mu$m, fairly round.  From the bottom every other layer
of 2 cm is filled with pink sand/yellow glass and gray sand/black glass
until layer 3. The cell is sheared exactly 2 cm, which means that opposite
colors tell the shear zone's position. The material is vacuumed away and
photos made layer by layer. At layer boundaries the colors change,
in-between the contrast is good for at least 1.5 cm.

Except at the ends, the system is invariant in the vertical direction,
reducing the two-dimensional problem of Eq. (\ref{eq:surfint}) to the one of
finding the easiest path for the shear zone between its two end
points. This line is not the shortest path but the path that minimizes
dissipation. The problem is exactly the same as in geometrical optics,
where Fermat's principle of least time implies Snell's refraction law. This
gives
\begin{equation}
\label{eq:snell}
  \mu_1 \sin{\theta_1} = \mu_2 \sin{\theta_2}
\end{equation}
at the material boundary, where the index of refraction from optics is
replaced by an effective material index.

\begin{figure}[]
\includegraphics[width=0.95\columnwidth]{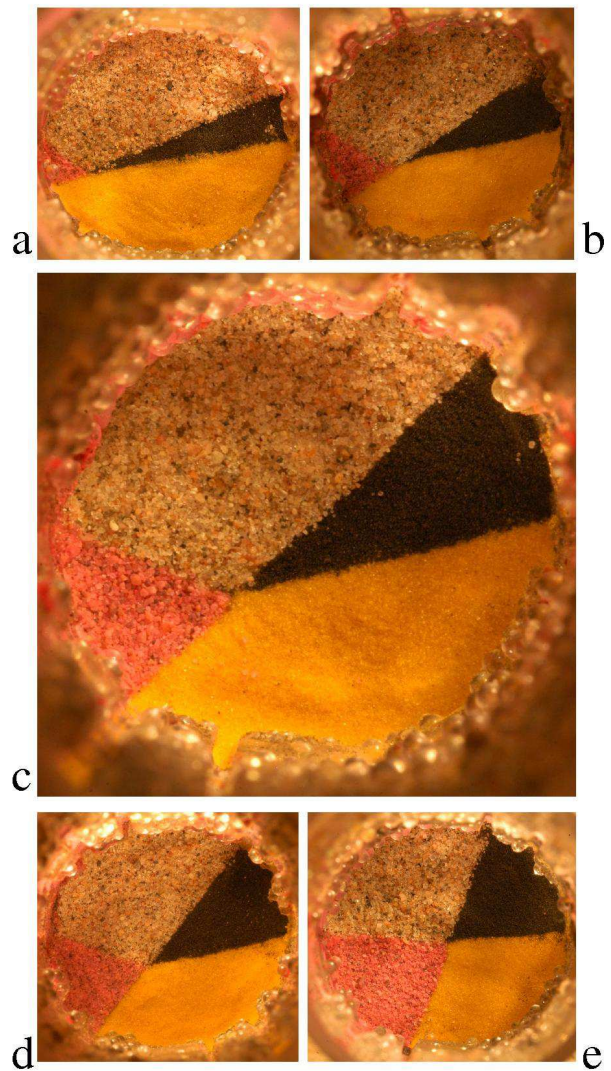}
\caption{\label{fig:expimg}
Refraction at the material boundary. Its orientation $\phi$ is: (a)
22$^{\circ}$, (b) 36$^{\circ}$, (c) 45$^{\circ}$, (d) 54$^{\circ}$, and (e)
67$^{\circ}$. The images are from layer 7 in (b-d) and layer 5 in (a) and
(e). The shear zone clearly deviates from the shortest path in all cases.
}
\end{figure}

We have performed the experiment with five different angles of $\phi$:
22$^{\circ}$, 36$^{\circ}$, 45$^{\circ}$, 54$^{\circ}$, and 67$^{\circ}$; shown
as (a) to (e) in Figure \ref{fig:expimg}\cite{footnote1}. The shear zone is
the line between pink and gray sand and the line between yellow and black glass
beads. In each of the materials we observe that the shear zone is a
straight line, supporting the assumption that one may assign an effective
friction and a material index to each material. Second the two line
segments meet in a point at the material boundary, and is thus refracted as
predicted by Snell's law.

\begin{figure}[]
\includegraphics[width=0.95\columnwidth]{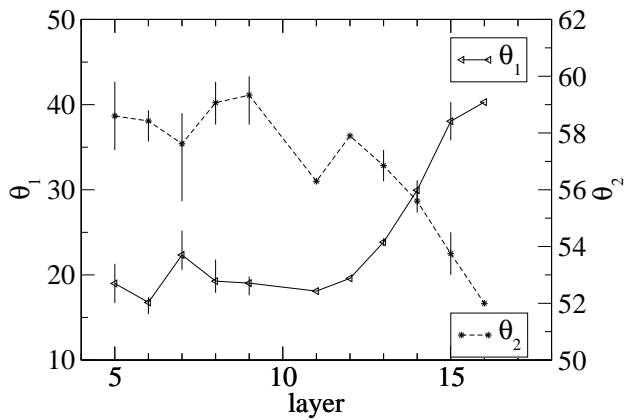}
\caption{\label{fig:angles}
The angles of incidence (defined in Figure \ref{fig:setup}) are shown as a
function of depth for the $\phi{=}45^\circ$-experiment. Here, about seven
layers show roughly the same behavior. There is a cross-over towards the
bottom to meet with the imposed boundary conditions. The error bars reflect
the variation within each layer (no error bar means that only one image was
obtained in that layer).}
\end{figure}

Qualitatively the effect is very clear. However, one should keep in mind
that the material index can not be expected to be history independent. The
preparation involves some gentle tapping to let the material settle; the
settling or density will vary to some degree between and within
experiments.  Furthermore, the system is subjected to gravity, which may
give different packing as a function of depth in the shear cell. For the
experiment in Figure \ref{fig:expimg}(c) ($\phi{=}45^{\circ}$) we show the
measured angles of incidence as a function of depth, see Figure
\ref{fig:angles}. There are some fluctuations due to inhomogeneity in the
preparation.  We observed that small slips could occur at the material
boundary, which locally perturb the shear zone. At the top, a few layers
are influenced by the loose mass sliding over, and at the bottom a straight
shear zone is imposed by the setup. Both effects are end effects, and in a
large portion in the middle of the cell the shear zone is stable.

\begin{figure}[]
\includegraphics[width=0.95\columnwidth]{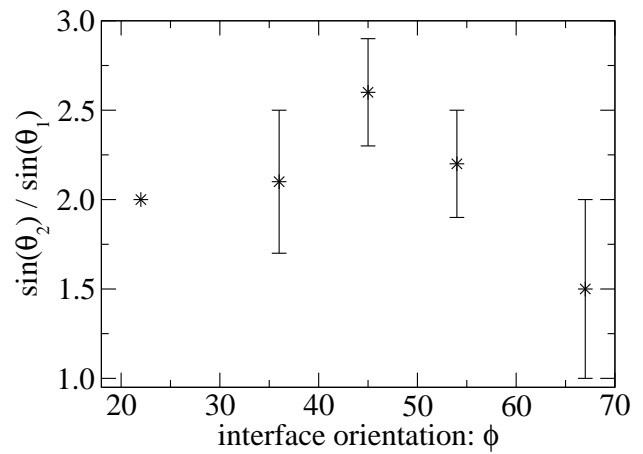}
\caption{\label{fig:ratios}
For each experiment the material index ratio is estimated. For the
$\phi{=}22^\circ$-experiment there was problems with slip at the material
boundary and the value is based on one layer only. Otherwise the error bars
reflect the fluctuations of the data.}
\end{figure}

From each of the five experiments the material index ratio is determined,
see Figure \ref{fig:ratios}. For each angle an average is made over the
available levels, that is to say excluding levels where slip or other
irregular behavior was seen. In numerical simulations where the properties
of each material are perfectly controllable, the index ratio was tuned and
in that way the measured index ratio ($\mu_1 / \mu_2 = \sin\theta_2 /
\sin\theta_1$) was checked against an expected value\cite{unger07}. In
these experiments the expected index ratio is an unknown, not only
dependent on the material, but also on the preparation -- the history of
the system. We measure the index ratio to be between 1.5 to 2.6. The shear
resistance of the sand and the glass was tested separately, to get an order
of magnitude estimate of their index ratio. A flat disk on a rod was
submerged in a beaker with the material, and upon pulling it out the
required force was measured. Loosely packed, the sand to glass friction
ratio was 1.5, whereas material compacted by tapping gave the ratio
2.5. Firstly, this tells that the index ratios found from Snell's law are
reasonable. Secondly, the index ratio depends on the preparation to a large
degree. A natural interpretation is that the irregular shape of sand makes
its index increase when the material is compacted. As a consequence one
should understand the $\mu_{1,2}$ in Eq. (\ref{eq:snell}) to include the
preparation history in order for the principle to hold.  Nevertheless,
Snell's law as a manifestation of the minimum dissipation principle for a
well prepared layered sample is shown to be valid. As such it confirms the
validity of the principle itself. It is clear from the experiments that
increasing inhomogeneity of the packing soon ruins the applicability of
Snell's law, however, this does not invalidate the general principle. 

\begin{figure}[th!]
\includegraphics[width=0.95\columnwidth]{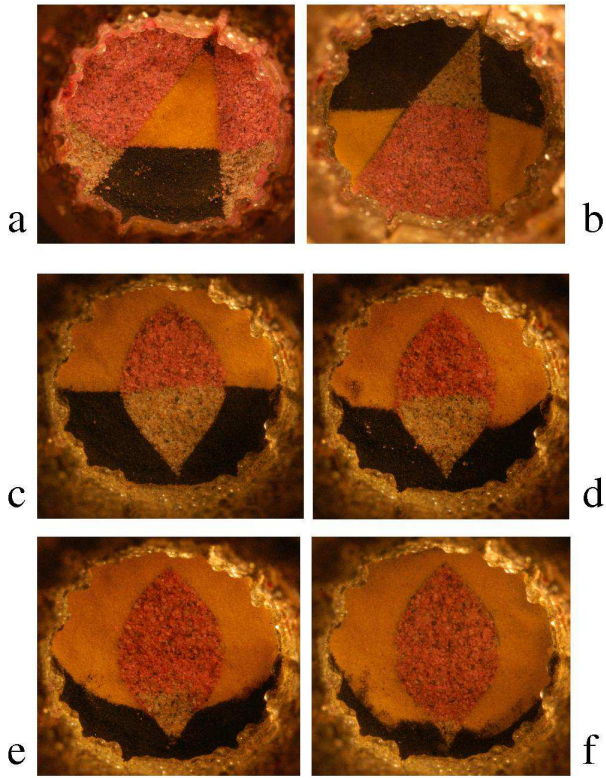}
\caption{\label{fig:lens}
Refraction at two material boundaries. (a) A wedge of glass beads in sand
and (b) a wedge of sand between glass beads. For both wedges the shear zone
was sharply defined. (c-f) A ``granular lens'' of sand showed a very wide
shear zone. The pictures are all from within layer ten at the
positions(measured from the top of the layer) : (c) 5 mm, (d) 9 mm, (e) 13
mm, and (f) 18 mm. This apparent shift of the shear zone within a layer
actually means that there is bulk flow over nearly half the tube width.
}
\end{figure}
Moving on from the single refraction setup, we demonstrate the effect of
refraction at two boundaries by looking at a wedge geometry, see Figure
\ref{fig:lens}(a-b). The shear zone is refracted twice as Snell's law
predicts. We observed that this shear zone is narrow and invariant with
depth in the tube. This is as expected when there is a clear minimum of the
functional in Eq. (\ref{eq:surfint}). Consider instead of the wedge shape a
granular lens centered in the cell. A convex lens of sand between glass
beads will, when properly shaped according to the material index ratio,
make a continuous range of shear zones which fulfill Snell's law. They have
equal dissipation rate, thus, the minimum dissipation principle as
formulated in Eq. (\ref{eq:surfint}) fails to select one of the shear
zones.

We have done the shear experiment with a granular lens that, save for the
difficulties in preparation, is designed after an index ratio of 2.0. The
result is shown in Figure \ref{fig:lens}(c-f).  The shown images are all
from within a single layer. As opposed to a narrow zone that makes all cuts
within a layer look identical, these images show a strong shift of the
apparent position of the shear zone. That is the signature of a very wide
shear zone and bulk flow. The fact that the bulk flow did not spread out
over the entire lens may be because of imperfections in the lens
shape. Note that a successful experiment requires both the shape of the lens
and the material index ratio to be accurately prepared. We found that small
variations in lens thickness or filling could ruin the lens effect.

The granular lens experiment shows that in that case the narrow shear band
assumption that allowed us to go from the volume integral in
Eq. (\ref{eq:volint}) to the surface integral of Eq. (\ref{eq:surfint}) is
invalid. Snell's law is not directly invalidated in the case of the lens
in the sense that the shear to some approximation is distributed over a
range of allowed shear zones. However, it does not provide a way to tell 
how the shear would be distributed among these. Due to experimental
difficulties it would be very interesting to approach this problem
numerically.

In this study we have applied the principle of minimum dissipation to a
particular problem. It is reasonable to think that there must exist some
range of validity, and we speculate that the following circumstances are
essential. The input energy is dissipated almost immediately, there is
energy balance, and the process is in steady-state in this sense. In order
for the system to be able to select a global minimum, information about the
driving force has to spread across the system at a time scale faster than
the change in driving force. In practice, the imposed strain from the
moving outer wall is transferred from grain to grain throughout the shear
cell, and this sets, at least in principle, a limit on the the global
strain rate.

In conclusion we present shear experiments with two materials in contact,
separated by a plane material boundary. Shearing through both materials we
obtain a narrow and well-defined shear zone. The shear zone is shown to
refract at the material boundary. By assigning a \emph{material index} to
each material, the refraction is found to obey Snell's law. Thereby we
confirm the minimum dissipiation principle in Eq. (\ref{eq:surfint}).

We seek to further investigate the range of validity of the extremum
principle. Inspired by optics we design a \emph{granular lens}. When
properly shaped, a continuous range of shear zones obey Snell's law and
have the same total dissipation rate. Shearing through the lens is shown to
produce a very wide shear zone. The minimum principle in the form of
Eq. (\ref{eq:volint}) or (\ref{eq:surfint}) does not provide a
way to tell how shear will be distributed between possible shear zones.
If investigated with sufficient accuracy, the granular lens may be used to
investigate the validity of the minimum principle or corrections to the
dissipation integral. We believe that a numeric approach to this will be
very valuable.

\begin{acknowledgments}
We thank Arvid Andreassen and Ken Tore Tallakstad for assistance with the
experimental setup. The experiments were performed in the laboratory of
Knut J{\o}rgen M{\aa}l{\o}y. This work was supported by the Norwegian
Research Council.
\end{acknowledgments}


\begin{thebibliography}{17}
\bibitem{footnote1}
  The cited angles are nominal and we attempted to make these angles. Small
  deviations are inevitable, but all measurements of angles
  are made relative to the material boundary so that these nominal angles do not
  enter explicitly in the results.
\end{thebibliography}

\end{document}